# Generation of concentration density maxima of small dispersive coal dust particles in horizontal iodine air filter at air-dust aerosol blow

I. M. Neklyudov, O. P. Ledenyov, L. I. Fedorova, P. Ya. Poltinin

*Solid State Physics, Materials Science and Technologies Institute*
*National Scientific Centre Kharkov Institute of Physics and Technology,*
*Academicheskaya 1, Kharkov 61108, Ukraine.*

The spatial distributions of the small dispersive coal dust particles with the sizes of below *1 μm* and below *10 μm* in the granular filtering medium with the cylindrical coal granules in the absorber in the horizontal iodine air filter during its long term operation at the nuclear power plant are researched. It is shown that the concentration density maxima of the small dispersive coal dust particles appear in the granular filtering medium with the cylindrical coal absorbent granules in the horizontal iodine air filter at an action by the air-dust aerosol blow. The comparison of the measured aerodynamic resistances of the horizontal and vertical iodine air filters is conducted. The main conclusion is that the magnitude of the aerodynamic resistance of the horizontal iodine air filters is much smaller in comparison with the magnitude of the aerodynamic resistance of the vertical iodine air filters at the same loads of the air – dust aerosol volumes. It is explained that the direction of the air - dust aerosol blow and the direction of the gravitation force in the horizontal iodine air filter are orthogonal, hence the effective accumulation of the small dispersive coal dust particles takes place at the bottom of absorber in the horizontal iodine air filter. It is found that the air – dust aerosol stream flow in the horizontal iodine air filter is not limited by the appearing structures, made of the precipitated small dispersive coal dust particles, in distinction from the vertical iodine air filter, in the process of long term operation of the iodine air filters at the nuclear power plant.

## Introduction

The precise characterization of the process of the air-dust aerosols propagation in the dispersive granular filtering medium can help to improve the designs of the iodine air filters (*IAF*), which are used for the radioactive chemical elements and isotopes emission reduction at the nuclear power plants (*NPP*). The problem is that the magnitude of the *IAF's* aerodynamic resistance increases nonlinearly at the reach of the critical magnitude of the small dispersive coal dust particles concentration at the narrow sub-surface layer in the granular filtering medium (*GFM*) with the cylindrical coal granules in the absorber in the *IAF* [1].

In this research, the methods of experimental modeling are applied to investigate the physical features of the dependences of the magnitude of aerodynamic resistance on the mass share of the small dispersive coal dust particles load in the horizontal *IAF* at the air-dust aerosol blow process. The spatial distributions of the small dispersive coal dust particles along the *IAF's* absorber length and across the *IAF's* absorber cross-section are also researched in details.

## Methodology of experiment to research small dispersive coal dust particles maxima generation in granular filtering medium with cylindrical coal granules in absorber in iodine air filter model

The *IAF's* models have been developed in the frames of the nuclear science and technology fundamental research program at the *NSC KIPT*. The *IAF's* model has a cylindrical shape with the diameter of *10 cm* and the length of the granular filtering medium (*GFM*) of *30 cm*, which corresponds to the thickness of the granular filtering medium (*GFM*) with the cylindrical coal granules in the absorber in the *IAF* at the *NPP* in Fig.1 (a). The model parameters are selected in such a way that the mean magnitude of the air stream flow velocity in the *IAF's* model is comparable to the magnitude of the air stream velocity in the real *IAF* at the *NPP*. At these conditions, the aerodynamic resistances of both the *IAF's* model and the real *IAF* coincide. The *IAF's* model includes the ten metallic containers in which the layers of the granular filtering medium with the cylindrical coal granules are fixed by the grids with the big cells. Every container was divided on the four demountable segments by the special grids as shown in Fig. 1 (b). The segments with the identical numbers create the four layers of absorber along the *IAF's* model. The segments of containers were filled with the specially selected big cylindrical coal granules of the absorber of the type of *CKT-3* (the coal granule's diameter is *1,8 mm*, the length is *3,2 mm*). In the front of the *IAF's* model, the container (a dust source) with the mix of the small dispersive coal dust particles and the cylindrical coal granules (not more than *1,5 %* of the small dispersive coal dust particles from the total mass of absorbent) was placed. The small dispersive coal dust particles with the dimensions below *1 μm* were synthesized at the special centrifugal mill, operating at the high velocities of rotation, in the process of forced crushing of the cylindrical coal adsorbent granules of the type of *CKT-3*. The mix of the small dispersive coal dust particles and the cylindrical coal granules in the dust source container was reloaded to preserve the same starting conditions in every subsequent experiment.

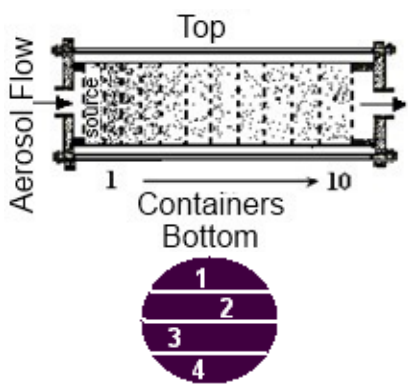

*Fig. 1. (a) Scheme of absorber with containers in iodine air filter (IAF) model;*
*(b) Scheme of cross-section of container with 1-4 segments, shown from top to bottom.*

In the case of every next experiment, the mass of the small dispersive coal dust particles, displaced into the *IAF*, was determined as a difference between the mass of dust source container after the experiment completion and the mass of dust source container before the experiment beginning. After the completion of a full research program cycle, the values of masses of the small dispersive coal dust particles, precipitated in every segment of all the containers in the absorber in the *IAF's* model, were precisely measured.

In the experimental research, the following technical variables are used:
$M_o$ is the mass of absorbent with the cylindrical coal adsorbent granules in the *IAF's* model;
$M^j_i$ is the mass of absorbent in the *j*-segment (*j* changes from *1* to *4*) in the container no. *i* (*i* changes from *1* to *10*) in the *IAF's* model;
$m_o$ is the total mass of the small dispersive coal dust particles, which was displaced into the absorber from the dust source container in the *IAF's* model at the experiment completion;

$m^j_i$ is the mass of the small dispersive coal dust particles, which was accumulated in the *j*-segment (*j* changes from *1* to *4*) in the container no. *i* (*i* changes from *1* to *10*) in the *IAF's* model at the experiment completion;
$h$ is the length of granular filtering layer in the *IAF's* model;
$\Delta P$ is the difference of pressures at the input and output in the *IAF's* model, measured by the water manometer.

## Experimental measurements results on spatial distribution of small dispersive coal dust particles along length of granular filtering medium layers in absorber in iodine air filter's model

The mean values of the mass shares of the small dispersive coal dust particles, precipitated in every of the four segments in the *IAF's* model at the completion of experimental researches, $m^j_i/(M^j_i + m^j_i)$, were obtained for all the ten containers. The graphics of distribution of the small dispersive coal dust particles across the cross-section of the containers *1-10* are presented in Fig. 2. The experimental results show that there is a transposition of the small dispersive coal dust particles to the bottom layers of absorbent from the first upper layer of absorbent at the sub-surface granular filtering medium in the *IAF's* model. At the end of experiment, the mass share of the small dispersive coal dust particles in the segment no. *1*, which is situated nearby the dust source container, is almost in the two times smaller, comparing to the segment no. *2* in the researched *IAF's* model as shown in Curve 1 in Fig. 2.

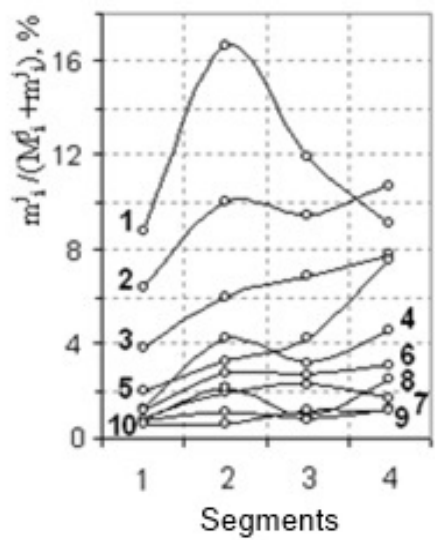

*Fig. 2. Distribution of mass share of small dispersive coal dust particles fraction across cross-section no. j (j changes from 1 to 4) in container no. i (i changes from 1 to 10) in iodine air filter (IAF) model.*

Using the data in Fig. 2, the curves to characterize the distribution of mass shares of the small dispersive coal dust particles along the length *h* of the four adsorbent layers are drawn in Fig. 3 (a). As it can be seen from the curves, in the sub-surface absorbent layers, there are the clearly visible dense regions with the small dispersive coal dust particles fraction with the mass shares equal to: *8,8 %; 16,6 %; 11,9 %;* and *9,1 %* in the segments no. *1-4* correspondingly. Then, there is a sharp decrease of the mass shares of the small dispersive coal dust particles so that the quantity of the small dispersive coal dust particles is decreased in *7,0; 4,0; 3,7;* and *2* times in the segments no. *1-4* on the length of *6 cm* approximately. Further, there is a smooth reduction of the mass share of the small dispersive coal dust particles in all the four layers down to *0,7 %* in the cases of the *1-2$^{nd}$* layers and *1,2 %* in the cases of the *3 - 4$^{th}$* layers at the output of the *IAF's* model. At all the *IAF's* length, the values of mass shares of the small dispersive coal dust particles in the segments of the first upper absorbent layer (see Curve 1 in Fig 3 (a)) are much smaller than in the segments of the bottom absorbent layers. Starting with the second container, the quantity of the small dispersive coal dust particles in the segments of the fourth bottom layer (Curve 4 in Fig. 3 (a)) exceeds the values in the segments of overlying layers. The analysis of obtained data shows that, at the action by the gravitation force, there is a displacement of the incoming small dispersive coal dust particles toward the direction of absorbent layer's bottom.

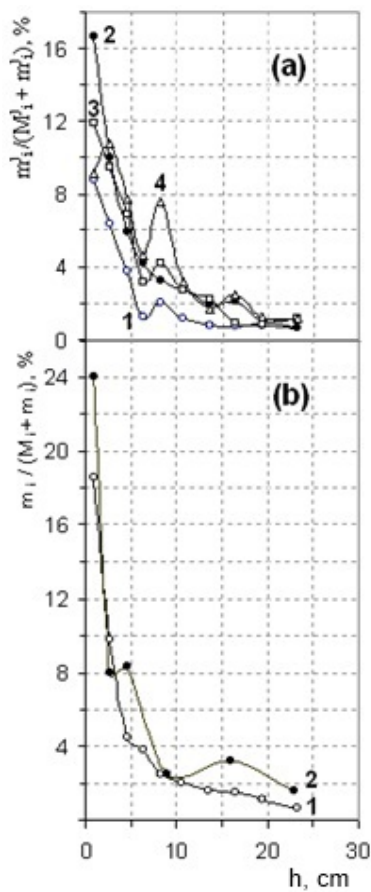

***Fig. 3.*** *a) Distribution of small dispersive coal dust particles fraction along length of 1-4$^{th}$ layers of adsorbent (j is from 1 to 4) in horizontal iodine air filter at end of all series of experiments; size of small dispersive coal dust particles is less than 1 μm; i is from 1 to 10.*
***b)*** *Distribution of small dispersive coal dust particles fraction along length of absorbent layer in vertical iodine air filter at end of all series of experiments; size of small dispersive coal dust particles is (1) less than 1 μm; (2) less than 10 μm; i is from 1 to 10.*

In Fig. 3 (b), the distributions of the small dispersive coal dust particles fraction along the length of absorber in the vertical *IAF* in the cases of: *(1)* small size dust and *(2)* big size dust fractions are shown [1-2]. As it follows from the comparative analysis data in Figs. 3 (a, b), in the case of the small size dust fraction in the absorber in the horizontal *IAF*, there are the maxima of the small dispersive coal dust particles masses concentration density in the granular filtering medium with cylindrical coal granules, which are positioned on the distances of *3 cm* – for the *4$^{th}$* layer, *8 cm* for the *1$^{st}$*, *3$^{rd}$* and *4$^{th}$* layers; *16 cm* – for the *2$^{nd}$* and *4$^{th}$* layers. This is a main distinction of the horizontal *IAF* from the vertical *IAF*. This type of the maxima of the small dispersive coal dust particles masses concentration density was early observed in the vertical *IAF*, when the big size dust fraction propagated in the vertical *IAF* (Curve 2 in Fig. 3(b)). According to the diffusion model in [3], in the case of the considered dependences of the small dispersive coal dust particles masses concentration density on the distance *x* along the length of absorber in the *IAF*, it is possible to introduce the general formula to describe a main peak of the small dispersive coal dust particles masses concentration density, which appears due to the process of structurization of the small dispersive coal dust particles of various sizes in a close proximity to the absorber's input surface and it can be described by the expression:

$$C(z) = C_{Z=0}[1 - erfz] = C_{Z=0} erfcz, \quad (1)$$

where *erfz* is the "*Gauss* integral of errors", and in the considered case $z = \dfrac{3^{1/2}(x-1)}{2(Dt)^{1/2}}$; *D* is the general diffusion coefficient of the small dispersive coal dust particles; *t* is the time of the air – dust aerosol blow ($t = 7,02 \times 10^4$ *sec*). The value of diffusion coefficient is equal to $D = 8 \times 10^{-8}$ *cm$^2$/sec* in the case of the initial part of the Curves 1 and 2 in Fig. 3 (a).

The solution of equation for the maxima of the small dispersive coal dust particles masses concentration density in the core of the *IAF*, which are connected with the creation of the small dispersive coal dust particles clots of particular size, can be written in the form of the *Gauss* distribution, which has a trend to be displaced in the depth of the *IAF* over the certain time period [3]:

$$C(x) \propto \left\{\dfrac{Q_0}{(\pi D_i t)^{1/2}}\right\} \exp\left(\dfrac{(x-x_0)^2}{4 D_i t}\right), \quad (2)$$

where $Q_0 = \int C(x, t) dx$; $x_0$ is the position of center of peak, $x_0 \sim V_i t$; $V_i$ is the average velocity of peak's movement at the displacement; the values with the index $i$ relate to the small dispersive coal dust particles of certain size, which are characterized by the index $i$.

The appearance of the small dispersive coal dust particles clots is due to the existing dependence of the small dispersive coal dust particles transposition velocity on the small dispersive coal dust particles size, and is facilitated by the circumstance that the exchange of impulses at the collisions among the small dispersive coal dust particles occurs most intensively in the case, when the small dispersive coal dust particles have the similar masses, and hence, the similar sizes.

In Fig. 4, the dependencies of distributions of the small dispersive coal dust particles masses fraction along the length of the $4^{th}$ absorbent layer at the bottom of the *IAF's* absorber are shown: *(1)* the dependence is a smoothed experimental curve; *(2)* the dependence is a computed curve for a monotonous part of the distribution; *(3)* the dependence is a computed curve for the average dust peak.

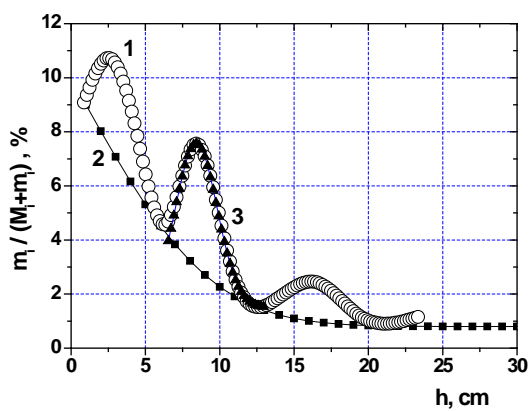

***Fig. 4.*** *Distribution of small dispersive coal dust particles masses fraction along length of $4^{th}$ absorbent layer at bottom of IAF's absorber: (1) dependence is smoothed experimental curve; (2) dependence is computed curve for monotonous part of distribution; (3) dependence is computed curve for average dust peak.*

The small dispersive coal dust particles with the smaller sizes reach the highest movement velocities in the absorber; and it makes sense to note that these particles create the clots, which are situated far away from the input surface of absorber in the *IAF*. These peaks of the small dispersive coal dust particles masses concentration density are clearly visible on the Curve 1 in Fig. 4, which characterizes to the absorbent layer at the bottom of the *IAF's* absorber as shown in Fig. 3 (a). The Curve 2 in Fig. 4 corresponds to the calculation, which is made, using the equation *(1)* with the diffusion coefficient, $D = 2,65 \times 10^{-8}$ $cm^2/sec$. The Curve 3 in Fig. 4 is obtained as a result of the calculation with the use of the formula *(2)* with the diffusion coefficient of the small dispersive coal dust particles in the given peak: $D = 9,72 \times 10^{-6}$ $cm^2/sec$. The average velocity of transposition of the small dispersive coal dust particles masses concentration density peak is equal to: $v = 1,17 \times 10^{-4}$ $cm/sec$. In the bottom segment of the absorber, a biggest part of the small dispersive coal dust particles, which were precipitated at an action by the influence by the gravitational force in the process of experiment, is situated. It is a characteristic fact that there is a considerable number of the small dispersive coal dust particles maxima in the bottom segment of the absorber (Curve 1 in Fig. 4). This result can only be explained by the integrated action by the gravitation force on the small dispersive coal dust particles, resulting in the small dispersive coal dust particles displacement to the bottom absorbent layer in the absorber in the *IAF's* model.

The mass share of the small dispersive coal dust particles, which was jettisoned by the air stream outside the *IAF*, *Δm*, was calculated as a difference between the total mass of the entered small dispersive coal dust particles in the absorber and the total mass of the precipitated small dispersive coal dust particles in all the segments of absorber. In the researched case, the mass share of the small dispersive coal dust particles, which was jettisoned by the air stream outside the *IAF*, is *Δm = 49,3 %*.

### Influence by spatial distribution of small dispersive coal dust particles in granular filtering medium with cylindrical coal granules in absorber on aerodynamic resistance of horizontal iodine air filter model

The dependence of the *IAF's* aerodynamic resistance on the volume of the air – dust air stream flow for the various values of the entered small dispersive coal dust particles masses was measured in the experiment (the selected results are shown on the Curves 1-9 in Fig. 5). The selected data for the vertical *IAF* with the accumulated small dispersive coal dust particles fraction is presented on the Curves **1** - **6** in Fig. 5.

The comparative analysis of the obtained results for the vertical and horizontal *IAFs*, through which the air – dust aerosol with the small dispersive coal dust particles with the size of smaller than *1 μm* was blown, confirms the fact that the aerodynamic resistances of the two *IAFs* increase synchronously in the beginning of the experiments, when the $m_o / (M_o+m_o)$ values increase up to *2 %*. Then, at the equal values of the introduced small dispersive coal dust particles mass fraction in the vertical and horizontal *IAFs*, the magnitude of aerodynamic resistance in the horizontal *IAF* starts to increase faster and surpasses the magnitude of aerodynamic resistance in the vertical *IAF* on the approximate value of *10 – 8 %* (see the Curves 4 and **2**, 5 and **3**, 6 and **4**, 7 and **5**, 9 and **6** in Fig. 5).

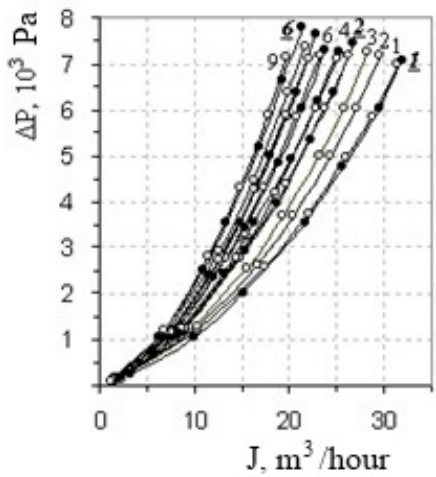

*Fig. 5. Dependences of absorber's aerodynamic resistances on volumetric air stream flow, depending on mass shares of introduced small dispersive coal dust particles fraction, in IAF, (%);*
*Sizes of small dispersive coal dust particles:*
*1) - Sizes of small dispersive coal dust particles are below1 μm in Curves 1- 9 and **1 - 6**;*
*2) ○ – horizontal IAF:1 (0), 2 (0,6), 3 (1,4), 4 (2,7), 5 (3,2), 6 (5,9), 7 (7,3), 8 (8,2), 9 (9,2);*
*3) ● – vertical IAF: **1** (0), **2** (4,3), **3** (5,6), **4** (6,4), **5** (7,9), **6** (9,3).*

In Fig. 6, the dependences of the aerodynamic resistance, normalized to the constant air stream flow: $J^* = 15\ m^3/cm$, at the magnitude of aerodynamic resistance: $\Delta P = 6000\ Pa$, on the relative mass share of the introduced small dispersive coal dust particles in IAF, $m_o / (M_o+m_o)$, are shown: *1, 3* – the horizontal *IAF*, *2, 4* – the vertical *IAF*. The sizes of the small dispersive coal dust particles are: *1 , 4* – below *10 μm*; *2, 3* - below *1 μm*.

On the Curve 3 in Fig. 6, the dependence of the aerodynamic resistance, $\Delta P^*$, which is normalized to the constant air stream flow $J^* = 15\ m^3/cm$ at $\Delta P = 6000\ Pa$, on the relative mass share of the introduced small dispersive coal dust particles in the IAF $m_o / (M_o+m_o)$ is presented in the case of the horizontal *IAF* with the small dispersive coal dust particles of the small sizes.

During the calculation of the aerodynamic resistance $\Delta P^*$ with the application of the empirical dependence $\Delta P = k \cdot J^{1,5}$ [1], the dependences $\Delta P(J)$, obtained in the present experiment, have been used (the selected data is presented on the Curves 1 - 9 in Fig. 5). In Fig. 6, the early obtained dependences of the aerodynamic resistance $\Delta P^*$ on the relative mass share of the small dispersive coal dust particles masses fraction $m_o / (M_o+m_o)$ in the cases of: *1)* the vertical *IAF* with the small dispersive coal dust particles of both the small sizes (Curve 2) and the big sizes (Curve 4); and *2)* the horizontal *IAF* with the small dispersive coal dust particles of the big sizes; which were continuously blown through the *IAF* (Curve 1) are shown [1, 2, 4].

In Fig. 6, the data allows us to compare the results of researches on the *IAFs* with both *1)* the various consistence of the small dispersive coal dust particles, which are blown through the granular filtering medium with cylindrical coal granules; and *2)* the various orientations of vectors of the physical forces, which act on the small dispersive coal dust particles in the *IAF*: *a)* the viscous capture force, which moves the small dispersive coal dust particles along the length of absorber; and *b)* the gravity force, which moves the small dispersive coal dust particles to the bottom of the *IAF's* absorber. In the vertical *IAF*, these physical forces are directed toward the same direction along the direction of the air – dust aerosol stream flow; in the horizontal *IAF*, these physical forces are directed at the right angle to each other.

The Curves 1 - 3 in Fig. 6 (Curves 1, 3 characterize the horizontal *IAF*, Curve 2 characterizes the vertical *IAF*, Curve 1 relates to the small dispersive coal dust particles of big sizes, Curves 2, 3 relate to the small dispersive coal dust particles of small sizes) are not quite different from each other. The dependence of the aerodynamic resistance on the relative mass share of the introduced small dispersive coal dust particles in the *IAF*, $\Delta P^* (m_o / (M_o+m_o))$, is close to the linear type of dependence with a little different angle of slope.

The Curve 4 in Fig. 6 (the vertical *IAF* with the small dispersive coal dust particles of big sizes) is close to the linear type of dependence approximately up to the saturation value of *7 %* of the relative mass share of the introduced small dispersive coal dust particles in the *IAF*, $m_o / (M_o+m_o)$. However, there is an exponential increase of the *IAF's* aerodynamic resistance at a further increase of the relative mass share of the introduced small dispersive coal dust particles in the *IAF*, $m_o / (M_o+m_o)$.

The physical features of the aerodynamic resistance physical behaviour in all the four cases in Fig. 6 can be explained, making a comparative analysis between the four curves in Fig. 6 and the characteristic distribution of the small dispersive coal dust particles along the *IAF* in Fig. 3.

In the horizontal *IAF* with the small dispersive coal dust particles of the big sizes [4], the small dispersive coal dust particles are displaced from the top to the bottom of the *IAF* at an action by the gravity force, excluding a possibility of the blocking structures creation, made of the small dispersive coal dust particles, in the close proximity to the input surface of the absorber in the *IAF* [3]. The small dispersive coal dust particles of the smaller sizes represent a certain part from the total mass share of the introduced small dispersive coal dust particles in the *IAF* at the experiment.

Thus, the small dispersive coal dust particles of the smaller sizes have a possibility to be fast enough moved along the *IAF* and a possibility to be jettisoned outside the *IAF*. The part of *34,6 %* of the introduced small dispersive coal dust particles *(Δm = 65,4 %)* is accumulated in the *IAF*. The aerodynamic resistance of the horizontal *IAF* with the small dispersive coal dust particles of the big sizes increases in *1,6* times and reaches *3270 Pa* at the small dispersive coal dust particles mass share load of *9,2 %* (Curve 1 in Fig. 6).

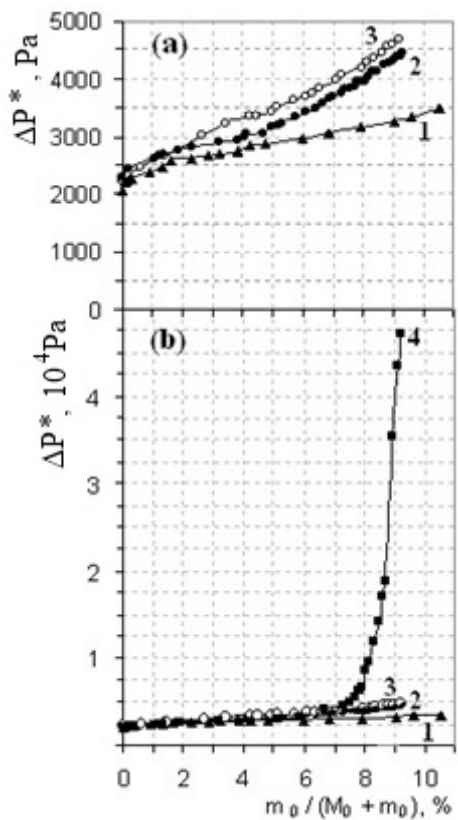

*Fig. 6. Dependences of aerodynamic resistance, normalized to constant air stream flow $J^* = 15\ m^3/cm$ at aerodynamic resistance $\Delta P = 6000\ Pa$, on the relative mass share of the introduced small dispersive coal dust particles in IAF, $m_o/(M_o+m_o)$:*
*1, 3 – horizontal IAF;*
*2, 4 – vertical IAF.*
*Sizes of small dispersive coal dust particles are:*
*1, 4 – below 10 $\mu m$; 2, 3 – below 1 $\mu m$.*

In the vertical *IAF* with the small dispersive coal dust particles of the small sizes, there is no the process of accumulation of the small dispersive coal dust particles (Curve 1 in Fig. 3(b)) [2]. The sub-surface layer of the granular filtering medium with the cylindrical coal granules continues to be well transparent for the transposition of the small dispersive coal dust particles masses at the experiment. The aerodynamic resistance of the vertical *IAF* with the small dispersive coal dust particles of the small sizes increases a bit faster, in *1,9* times, by the end of experiment, and equals to *4380 Pa* (Curve 2 in Fig. 6), $\Delta m = 61\ \%$.

Going from the completed experimental researches of the vertical *IAFs* with the small dispersive coal dust particles of the relatively small- and big- sizes, it is possible to make a conclusion that the use of the absorbent with the reinforced cylindrical coal granules can result in the much longer term of the *IAFs* operation, because the small dispersive coal dust particles, which can damage the absorber in the *IAF*, are originated in the process of the grinding of the cylindrical coal granules during the air blow.

During the air blow process, the dense soft condensed matter regions in the granular filtering medium with the cylindrical coal granules, $\Delta m = 49,3\ \%$, are formed in the researched case of the horizontal *IAF* at the air-dust aerosol blow with the small dispersive coal dust particles of the small sizes as well as in the case the vertical *IAF* at the air-dust aerosol blow with the small dispersive coal dust particles of the big sizes [1]. By the end of experiment, the aerodynamic resistance of the horizontal *IAF* increases in *2,1* times and reaches the magnitude of *4680 Pa* (Curve 3 in Fig. 6). In case of the air-dust aerosol blow through a layer of the granular filtering medium with the cylindrical coal granules in the absorber in the horizontal *IAF*, the small dispersive coal dust particles of the small sizes precipitate, collecting in the form of clots, in such a way that the mass share of the small dispersive coal dust particles is in *1,3* times bigger in the horizontal *IAF* than the mass share of the small dispersive coal dust particles in the vertical *IAF*, where the small dispersive coal dust particles maxima are not present [2]. After the completion of all the series of experiments with the application of the small dispersive coal dust particles of the very small sizes, it was found that the magnitude of the aerodynamic resistance of the horizontal *IAF* is in *1,1* times bigger that the magnitude of the aerodynamic resistance of the vertical *IAF*.

The magnitude of the aerodynamic resistance of the horizontal *IAF* is in *1,4* times bigger in the case, when the air –dust aerosol with the small dispersive coal dust particles of the small sizes is blown, comparing to the case, when the air – dust aerosol with the small dispersive coal dust particles of the big sizes is blown. It makes sense to note that a negligible part of the small dispersive coal dust particles of the small sizes is transported along the horizontal *IAF* only, comparing to the total mass share of the introduced small dispersive coal dust particles of the small sizes in the horizontal *IAF*. Therefore, the mass share of the accumulated small dispersive coal dust particles of the small sizes in the horizontal *IAF* is in *1,5* times bigger, comparing to the mass share of the accumulated small dispersive coal dust particles of the big sizes in the horizontal *IAF*. It is necessary to add the magnitude of the aerodynamic resistance is bigger, when the clots of small dispersive coal dust particles of the small sizes are present in the horizontal *IAF*; and the magnitude of the aerodynamic resistance is smaller, when the small dispersive coal dust particles of the small sizes are equally distributed in the horizontal *IAF*.

In the vertical *IAF*, the small dispersive coal dust particles fraction of big sizes mainly precipitated between the cylindrical coal granules of absorbent in the narrow layer with the thickness of *2 cm* in close proximity to the *IAF's* input surface, capturing the small dispersive coal dust particles of the smaller sizes in this granular filtering medium layer, and creating the almost monolithic layer with the critical mass of the small dispersive coal dust particles fraction of *18 %*, which was almost impenetrable for the air - dust aerosol propagation at the *IAF* operation [1]. A certain part of the small dispersive coal dust particles in the beginning of experiment especially, was accumulated in the form of the small dispersive coal dust particles clots, *Δm = 44,2 %*, in the absorber in the vertical *IAF*. As a result, the aerodynamic resistance increased sharply, in *23* times, up to the magnitude of *47060 Pa* (Curve 4 in Fig. 3).

**Conclusion**

The spatial distributions of the small dispersive coal dust particles with the sizes of below *1 μm* and below *10 μm* in the granular filtering medium with the cylindrical coal granules in the absorber in the horizontal iodine air filter during its long term operation at the nuclear power plant was researched. It is shown that the concentration density maxima of the small dispersive coal dust particles appear in the granular filtering medium with the cylindrical coal absorbent granules in the horizontal iodine air filter at an action by the air-dust aerosol blow. The comparison of the measured aerodynamic resistances of the horizontal and vertical iodine air filters is conducted. The main conclusion is that the magnitude of the aerodynamic resistance of the horizontal iodine air filters is much smaller (in *10* times) in comparison with the magnitude of the aerodynamic resistance of the vertical iodine air filters at the same loads of the air – dust aerosol volumes. It is explained that the direction of the air - dust aerosol blow and the direction of the gravitation force in the horizontal iodine air filter are orthogonal, hence the effective accumulation of the small dispersive coal dust particles takes place at the bottom of absorber in the horizontal iodine air filter. It is found that the air – dust aerosol stream flow in the horizontal iodine air filter is not limited by the appearing structures, made of the precipitated small dispersive coal dust particles, in distinction from the vertical iodine air filter, in the process of long term operation of the iodine air filters at the nuclear power plant. Therefore, the main conclusion is that the horizontal iodine air filter has a very small increase of the aerodynamic resistance, comparing to the vertical iodine air filter, at their long term operation at the nuclear power plant.

Authors are very grateful to a group of leading scientists from the *National Academy of Sciences in Ukraine* (*NASU*) for the numerous encouraging scientific discussions on the reported theoretical and experimental research results.

This innovative research is completed in the frames of the nuclear science and technology fundamental research program, facilitating the environment protection from the radioactive contamination, at the *National Scientific Centre Kharkov Institute of Physics and Technology* (*NSC KIPT*) in Kharkov in Ukraine. The research is funded by the *National Academy of Sciences in Ukraine* (*NASU*).

This research paper was published in the *Problems of Atomic Science and Technology* (*VANT*) in [6] in 2010.

*E-mail: ledenyov@kipt.kharkov.ua

\_\_\_\_\_\_\_\_\_\_\_\_\_\_

**1**. I. M. Neklyudov, L. I. Fedorova, P. Ya. Poltinin, L. V. Karnatsevich, Influence by physical features of accumulation of coal dust fraction in layer of adsorber on increase of aerodynamic resistance of coal iodine air filters in ventilation systems at nuclear power plant, *Problems of Atomic Science and Technology* (*VANT*), Series «*Physics of radiation damages and radiation materials*», no. 6 (84), pp. 65-70, ISSN 1562-6016, 2003.

**2**. I. M. Neklyudov, O. P. Ledenyov, L. I. Fedorova, P. Ya. Poltinin, Influence by small dispersive coal dust particles of different fractional consistence on characteristics of iodine air filter at nuclear power plant, *Problems of Atomic Science and Technology* (*VANT*), Series «*Physics of radiation damages and radiation materials*», no. 2 (93), pp. 104-107, ISSN 1562-6016, 2009.

**3**. O. P. Ledenyov, I. M. Neklyudov, P. Ya. Poltinin, L. I. Fedorova, Physical features of small disperse coal dust fraction transportation and structurization processes in iodine air filters of absorption type in ventilation systems at nuclear power plants, *Problems of Atomic Science and Technology* (*VANT*), Series «*Physics of radiation damages and radiation materials*», no. 3 (86), pp. 115-121, ISSN 1562-6016, 2005.

**4**. I. M. Neklyudov, L. I. Fedorova, P. Ya. Poltinin, O. P. Ledenyov, Features of coal dust dynamics at action of differently oriented forces in granular filtering medium, *Problems of Atomic Science and Technology* (*VANT*), Series «*Physics of radiation damages and radiation materials*», no. 6 (91), pp. 82-88, ISSN 1562-6016, 2007.

**5**. N. B. Ur'ev, Physical-chemical dynamics of dispersive systems, *Uspekhi Khimii*, vol. **79**, no. 1, pp. 39-62, 2004.

**6**. I. M. Neklyudov, O. P. Ledenyov, L. I. Fedorova, P. Ya. Poltinin, Generation of small dispersive coal dust particles concentration maximums in granular filtering medium with cylindrical coal absorbent granules in horizontal iodine air filter at action by air-dust aerosol, *Problems of Atomic Science and Technology* (*VANT*), Series «*Physics of radiation damages and radiation materials*», no. 6(96), pp. 67-72, ISSN 1562-6016, 2010 (in Russian)
http://vant.kipt.kharkov.ua/ARTICLE/VANT_2010_5/article_2010_5_96.pdf